\documentclass[preprint]{aastex631}

\usepackage{changepage}
\usepackage{graphicx} 

\begin{document}

\title{The Y Dwarf Companion to the White Dwarf WD 0806-66: Resolving the Discrepancy Between Atmospheric and Evolutionary Models}

\correspondingauthor{Sandy Leggett}
\email{sandy.leggett@noirlab.edu}

\author[0000-0002-3681-2989]{S. K. Leggett}
\affiliation{Gemini Observatory/NSF's NOIRLab, 670 N. A'ohoku Place, Hilo, HI 96720, USA}

\author[0000-0001-6172-3403]{Pascal Tremblin}
\affiliation{Universite Paris-Saclay, UVSQ, CNRS, CEA, Maison de la Simulation, 91191, Gif-sur-Yvette, France}

\begin{abstract}
{\it James Webb Space Telescope} near- and mid-infrared spectroscopy has been published by Voyer et al. (2025) and Lew et al. (2026) for the Y dwarf which is a distant companion to the white dwarf WD 0806-661 (Luhman et al. 2011).  This target is important because the distance and the age of the system are well constrained by the primary star.  Voyer et al. perform a retrieval analysis of the longer wavelength MIRI data, and Lew et al. perform retrieval and forward model grid analyses of the NIRSpec data. These studies produce different results, and both are discrepant with evolutionary model calculations based on the age of the system and the luminosity of the Y dwarf.  Here we confirm the luminosity of the Y dwarf, and update the age of the system to $1.6 ^{+0.6}_{-0.4}$~Gyr.
We compare the combined NIRSpec and MIRI dataset to synthetic spectra generated by ATMO 2020++ atmospheric models.  We find a good fit across the
entire observed spectral energy distribution, except at the shortest near-infrared wavelengths, with atmospheric parameters consistent with evolutionary models.  We find the Y dwarf to be slightly metal-poor, with an effective temperature of $357 \pm 3$~K, a radius of $1.08 \pm 0.02~R_{\rm Jupiter}$, and a mass of $7 \pm 1 ~M_{\rm Jupiter}$.
\end{abstract}


\bigskip
\section{Introduction} 

Brown dwarfs (BDs) are low-mass star-like objects which have insufficient mass for sustained energy production by nuclear fusion.  The coldest BDs are classified as Y dwarfs, and have $T_{\rm eff} < 500$~K \citep{Cushing_2011, Kirkpatrick_2021a}. The Y dwarfs have dynamic, turbulent, atmospheres which are similar to those of the solar system giant planets \citep[e.g.][]{Showman_2013}.  The coldest known example of this class is  WISE J085510.83$-$071442.5, an isolated BD with $T_{\rm eff} \approx 275$~K \citep{Luhman_2014,
Luhman_2024, Kuhnle_2024, Leggett_2025b}. 

BDs cool over periods of gigayears, and if age is not known there is a degeneracy between mass and effective temperature ($T_{\rm eff}$) --- a younger lower-mass BD can have the same $T_{\rm eff}$ as an older higher-mass BD \citep[e.g.][their Figure 13]{Marley_2021}.  Stellar systems which contain a BD, where age can be constrained, are therefore valuable for improving our understanding of BDs.  For the lowest mass BDs in stellar systems there is debate regarding whether such an object was formed like a star or like a planet \citep[e.g.][]{Bowler_2020}.  There is evidence that suggests 
an overlap in mass between low-mass objects forming via different channels \citep{Gilbert_2026}. Conventionally, ``planetary mass'' objects have masses below the deuterium burning limit of $\sim 13~M_{\rm Jupiter}$ \citep{Marley_2021}. However, there are isolated cold BDs in the solar neighborhood and in young clusters, which presumably formed like stars, that have masses $< 13~M_{\rm Jupiter}$ \citep[e.g.][]{Luhman_2024, Leggett_2025b}.  

This paper takes another look at the {\it James Webb Space Telescope (JWST)} data on a cold BD companion  to the white dwarf WD 0806-661 (also known as L 97-3).  Here we refer to the cold companion as a Y dwarf, although other studies have labeled it an exoplanet.   
Recent analyses of the {\it JWST} data by \citet[][hereafter Voy25]{Voyer_2025} and \citet[][hereafter Lew26]{Lew_2026}  have produced different values for the mass, temperature and radius of the Y dwarf, and these values in turn appear to disagree with evolutionary models of BDs (Lew26).  The specific observations can be accessed via \dataset[doi:10.17909/anme-j453]{\doi{10.17909/anme-j453}}
and \dataset[doi:10.17909/dacd-v990]{\doi{10.17909/dacd-v990}}.

In Section 2 we describe the system, in Section 3 we investigate fits to the {\it JWST} data, and in Section 4 the pressure-temperature relationship for the atmosphere of the Y dwarf is explored. Section 5 presents our conclusions. 

\bigskip
\section{The WD 0806-661 System}

\subsection{WD 0806-661}

WD 0806-661 was first presented as a white dwarf by \citet{Luyten_1949}.  Recent observations and model analyses show that it is a He-rich DQ white dwarf, with  $T_{\rm eff} = 10393$~K, a mass of $0.605 \pm 0.014~M_{\odot}$, and a cooling age of 0.63~Gyr \citep{OBrien_2024}.  The initial-final mass relation by \citet{ElBadry_2018} implies that the initial mass of the star was $2.06 ^{+0.30}_{-0.24}~M_{\odot}$.  Using the main sequence and post-main-sequence lifetime relationships from \citet{Mowlavi_2012}, the time spent prior to the white dwarf stage is 
$0.94 ^{+0.2}_{-0.4}$~Gyr. However, rapid rotation can increase the lifetime of a star on the main sequence by up to 30\% \citep{Nguyen_2022} and so we adopt a total age for the system of $1.6 ^{+0.6}_{-0.4}$~Gyr. 

\subsection{WD 0806-661B}

\subsubsection{Discovery}

The brown dwarf companion to WD 0806-661 was discovered by \citet{Luhman_2011}.  The BD is at a distance of 2500~au from the white dwarf.  This white dwarf-Y dwarf pair is the only confirmed example of its kind, however  \citet{Albert_MEAD_2025} announced a candidate Y dwarf companion to the white dwarf MEAD 62 (also known as  2MASS J09424023-4637176).   MEAD 62 is a magnetic DA white dwarf with a total age of $\approx 7.6$~Gyr \citep{OBrien_MEAD_WD}; if the companion is confirmed, the separation is 40~au, the temperature of the BD is $\approx 340$~K, and its mass is around the deuterium burning limit \citep{Albert_MEAD_2025}.

\subsubsection{{\it JWST} Observations and Previous Analyses}

Infrared photometry and spectroscopy for WD 0806-661B have been obtained using {\it JWST}. Lew26 present NIRCam F150W2, F200W, F356W, and F444W photometry, as well as a NIRSpec G395M spectrum spanning the wavelength range $3 < \lambda~\mu$m $< 5$. Voy25 present MIRI F1280W, F1500W, F1800W, and F2100W  photometry, as well as a MIRI LRS spectrum  spanning the wavelength range $5 < \lambda~\mu$m $< 12$.

Voy25 apply the  TauREx3 atmospheric retrieval code \citep{AlRef_2022} to the MIRI data to infer the properties of WD 0806-66B. Lew26 use PICASO \citep{Batalha_2019} for a retrieval analysis of the NIRSpec and NIRCam data, as well as a forward-model analysis using models from the ATMO2020++ \citep{Leggett_2023} and the Elf-Owl \citep{Mukherjee_2024} grids.  These two studies, Voy25 and Lew26, use different analysis tools, and they study different wavelength regions, probing somewhat different regions of the atmosphere.  While both find the atmosphere of WD 0806-66B to be slightly metal-poor, all other parameters -- $T_{\rm eff}$, radius ($R$), surface gravity ($g$), and mass ($M$) --  differ significantly (Lew26, their Table 5).

Another approach to determining the properties of this Y dwarf is to determine its bolometric luminosity, which, when combined with the age of the system and evolutionary models, constrains $T_{\rm eff}$, $R$, $g$, and $M$ \citep[][their Figures 10, 11]{Marley_2021}. Lew26 estimate the luminosity by combining the NIRCam, NIRSpec and MIRI data; the values of  $T_{\rm eff}$, $R$, and $g$ determined by their retrieval and forward model analyses disagree with the values inferred from the evolutionary models using the system's age and the Y dwarf's luminosity.  The Voy25 retrieval analysis values for $T_{\rm eff}$ and $R$ are consistent with the evolutionary models, however their values for $g$ and $M$ are not.

\bigskip
\section{A New Analysis of the {\it JWST} Data for WD 0806-661B}

\subsection{ATMO 2020++ Model Grid Comparison}

We compared the Voy25 and Lew26 data to synthetic spectra generated by the ATMO 2020++ models. In order to scale the modelled surface flux to the detected flux, the BD radius and distance are required. The distance to the WD0806-661 system, is well known \citep{Gaia_2020}. 
Voy25 and Lew26 allowed the radius of the cold Y dwarf companion to vary in their recent studies. Here, we make the assumption that substellar evolutionary models are more robust than the models of cold atmospheres, and we fix the radius to be the value calculated by the \citet{Marley_2021} evolutionary models for each \{$T_{\rm eff}$, $g$\} value in the ATMO 2020++ model atmosphere grid. 

The synthetic spectra were taken from the ATMO 2020++ phosphine-free grid
\href{https://www.erc-atmo.eu/}{ on the Opendata site}. 
The ATMO 2020++ models  adopt a diabatic profile with an effective adiabatic parameter of $\gamma  = 1.25$ (compared to the standard value of $\gamma \approx 1.4$).   The adiabat is modified at atmospheric pressures between 0.15 and 15 bar at log $g = 4.5$, which are scaled by $\times 10^{({\rm log}~g ~-~ 4.5)}$ at other surface gravities.  The modification is empirical, and was found to be necessary in order to reproduce the shape of the energy distribution from the near- to the mid-infrared for BDs with $T_{\rm eff} \lesssim 600$~K \citep{Leggett_2021}.  It is possible that convection is disrupted at these atmospheric depths by kinetic or chemical processes or both \citep{Tremblin_2015, Tremblin_2019, Tan_2021}.
 Out-of-equilibrium chemistry is used, with $K_{zz} = 10^5$ cm$^2$ s$^{-1}$ at log ~$g = 5.0$, scaled by
$\times 10^{2(5 - {\rm log} ~g)}$ at other surface gravities to reflect scale height changes with gravity \citep[e.g.][]{Marley_2015}. 
 The grid covers $250 \leq T_{\rm eff}$~K $\leq 1200$ (250~K, 275~K, 300~K, 350~K, 400~K, 450~K, 500K, then every 100~K to 1200~K), 
for three metallicities:  [m/H] $= -0.5, 0, +0.3$. The surface gravity $g$ ranges from log $g = 2.5$ to 
log $g = 5.5$ in steps of 0.5 dex. 
More information on the models is given in the README file on the web site, and in \cite{Leggett_2025a}.

\begin{figure}
\vskip -0.2in
\hskip 0.3in
\includegraphics[angle=-90, width = 7 in]
{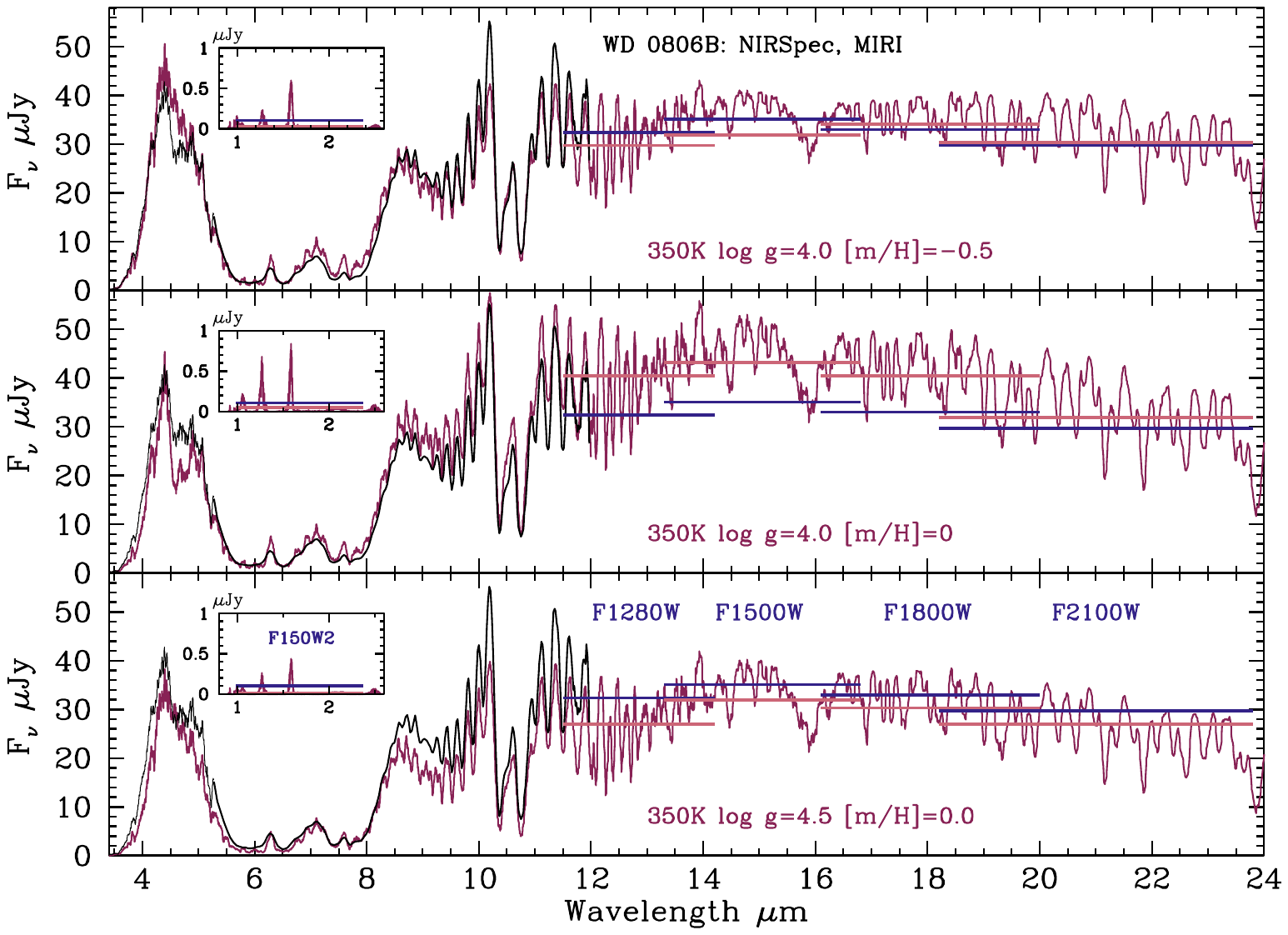}
\vskip -0.1in
\caption{Comparison of {\it JWST} spectra and photometry from Voy25 and Lew26 to ATMO 2020++ synthetic spectra and photometry \citep{Meisner_2023, Leggett_2023, Leggett_2024}, for WD 0806-661B. Black lines at $2.88 \leq \lambda~\mu$m $\leq 5.27$ are observed NIRSpec spectra, and black  lines at $5.27 \leq \lambda~\mu$m $\leq 11.99$ are observed MIRI spectra.  Indigo horizontal lines indicate observed photometric fluxes through the F150W2, F1280W, F1500W, and F2100W filters, as indicated in the legends in the bottom panel.  The purple lines are synthetic spectra generated by ATMO 2020++ atmospheric models with parameters as given in the legends in each panel.  Pink horizontal lines show the synthetic photometry for each filter and each model.  The model fluxes have not been scaled to match the observations, instead surface flux is converted to detected flux using the measured distance to the system of 19.23~pc \citep{Gaia_2020} and the 
value for the BD surface radius calculated by \citet{Marley_2021} evolutionary models for the \{$T_{\rm eff}$, $g$\} of each model.}
\end{figure}

Figure 1 compares the {\it JWST} data to the three best-matching ATMO 2020++ spectra.  The best fitting models were selected by eye, because the grid sampling is relatively coarse. All three models have $T_{\rm eff} = 350$~K. The 400~K synthetic spectra are $\sim 50\%$ too bright, and the 300~K spectra $\sim 50\%$ too faint, across the mid-infrared wavelength region (with radius fixed by evolutionary models).  All the 350~K models appear too faint at $1.0 \lesssim \lambda~\mu$m $\lesssim 2.0$, based on the observed F150W2 magnitude. This could be addressed by adjusting the adiabat of the atmosphere, making the deep atmospheric region (where the  near-infrared flux originates) warmer.  A {\it JWST} NIRSpec prism spectrum would be helpful to explore this further.

The log $g = 4.0$ spectra reproduce the flux peaks
at $\lambda \approx 4.5~\mu$m and $\lambda \approx 8.5~\mu$m better than the log $g = 4.5$ spectrum. The strength of the CO$_2$ absorption at $\lambda \approx 4.2~\mu$m and the CO absorption at $\lambda \approx 4.5~\mu$m is too strong in the solar metallicity model, and too weak in the [m/H] $= -0.5$ model, suggesting a metallicity between these two values.
The region around the strong NH$_3$ absorption  $\lambda \approx 10.5~\mu$m is well reproduced by both 350~K log $g = 4.0$ spectra, however the longer wavelength photometry is better reproduced by the more metal-poor spectrum. Metal-poor 2 Gyr-old stars do exist in the solar vicinity \citep[e.g.][their Figure 9]{Imig_2023} and so this result is plausible.  With a system component separation of 2500 au \citep{Luhman_2011}, atmospheric pollution of the BD during the evolution of the $2~M_{\odot}$ primary star would be unexpected, and so the BD atmosphere most likely remains in its original state.

\begin{figure}
\vskip -0.5in
\hskip 0.5in
\includegraphics[angle=0, width = 5.8 in]
{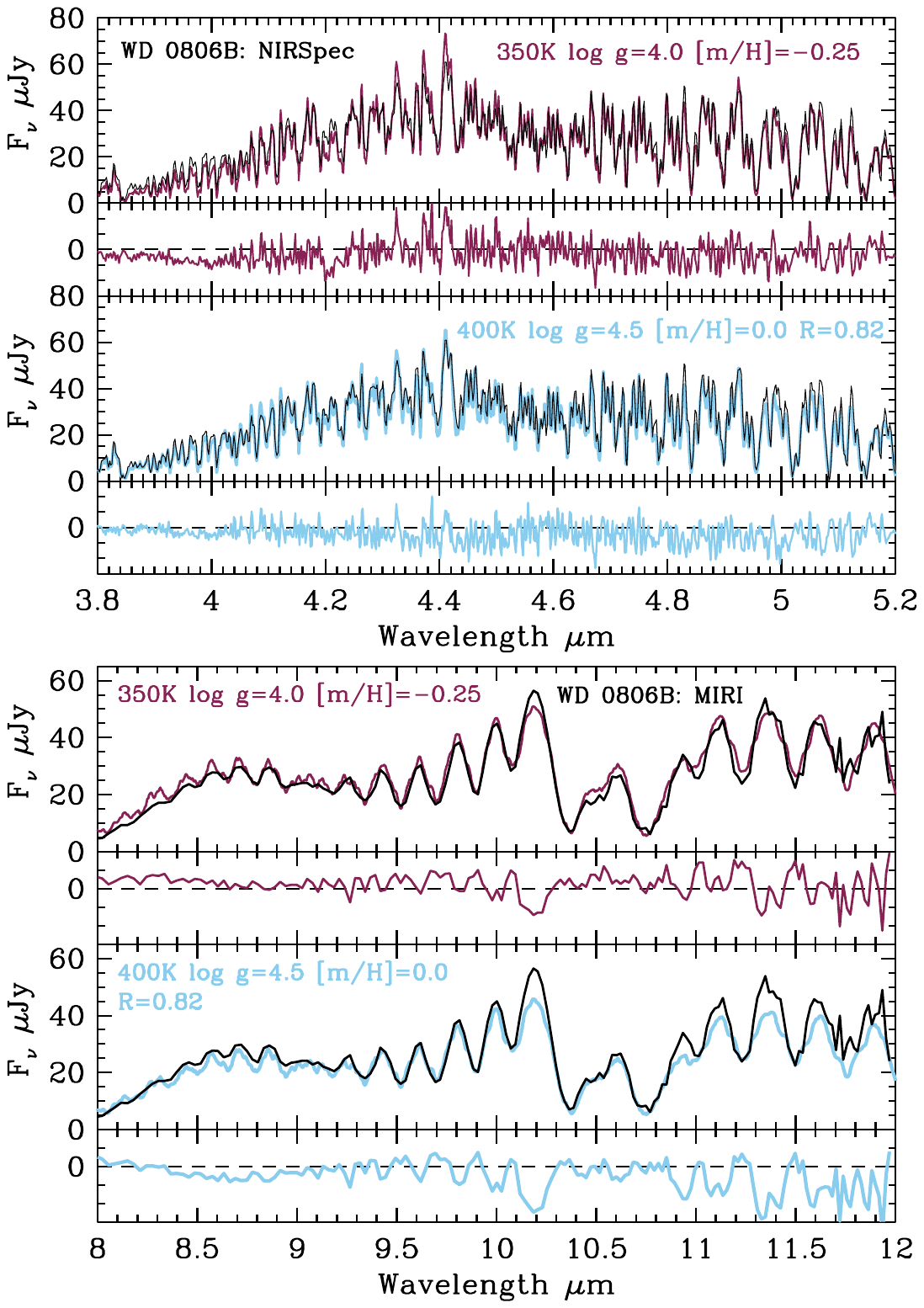}
\vskip -0.4in
\caption{Comparison of the Lew26 NIRSpec spectra (upper panels) and the Voy25 MIRI spectra (lower panels) to ATMO 2020++ synthetic spectra. The black line is the data and the colored lines are the model spectra. The purple line estimates the spectrum for $T_{\rm eff}=350$~K log $g = 4.0$ [m/H]$=-0.25$ by averaging the [m/H]$=0.0$ and [m/H]$=-0.5$ spectra, and uses the evolutionary model radius. The cyan line shows the ATMO 2020++ spectrum selected by Lew26 where the radius is allowed to float.
The difference $ model - observation $ is shown below each main panel, with a dashed line indicating zero.
}
\end{figure}

Lew26 also compare the NIRSpec spectrum for WD 0806-661B to ATMO 2020++ synthetic spectra. In their application, the radius of the BD was allowed to vary, and they find a best fit solution of $T_{\rm eff} = 400$~K, log $g = 4.5$, and [m/H] $= 0.0$.  The adopted radius is $0.82~ R_{\rm Jupiter}$, compared to the evolutionary model value for these parameters of $1.01 R_{\rm Jupiter}$, for a factor 0.66 reduction in flux.  

Figure 2 shows  ATMO 2020++ fits to the NIRSpec and MIRI spectra, using the average of the 350~K log $g = 4.0$ [m/H]$=0.0$ and
[m/H]$=-0.5$ spectra to simulate an [m/H]$=-0.25$ spectrum,
and the Lew26 fit using the floating radius.  While the ATMO 2020++ model adopted by Lew26 reproduces the NIRSpec data well, it is too faint at the MIRI wavelengths, by around 25\% (which extends to the longer wavelength photometry).  The PICASO retrieval fit to the NIRSpec data by Lew26 also produces a longer wavelength spectrum that is fainter than observed, by $~30\%$ \citep[][their Figure 8]{Lew_2026}. The  rescaled $T_{\rm eff} = 400$~K log $g = 4.5$ spectrum selected by Lew26 produces a superior fit to the NIRSpec data  compared to our solution (Figure 2), however the 
$T_{\rm eff} = 350$~K log $g = 4.0$ spectrum produces a  NIRSpec fit which is almost as good, and produces a superior fit to the MIRI data.  
The average deviation from the NIRSpec observations is $8\%$ worse,
and from the MIRI observations $22\%$ better, for our selected model.
Moreover the deviation at the MIRI wavelengths does not show the slope of the warmer model (Figure 2), and the adopted $T_{\rm eff} = 350$~K log $g = 4.0$ [m/H] $\approx -0.25$ model results in
physical properties which are consistent with evolutionary models.

Table 1 gives the adopted parameters for WD 0806-661B based on this grid comparison, compared to those of Voy25 and Lew26 using the retrieval techniques.
Uncertainties in the ATMO 2020++ model estimates are constrained by the grid step size.

\begin{deluxetable*}{ccccc}[!hb]
\vskip 0.075in
\tablecaption{Estimated Parameters for WD 0806-661B}
\tablehead{
\colhead{Parameter} &  \multicolumn{2}{c}{This work} &  \colhead{Voyer+ 2025} &  \colhead{Lew+ 2026} \\
    &  \colhead{ATMO 2020++} & \colhead{Luminosity and Age} &
    \colhead{MIRI} & \colhead{NIRSpec, NIRCam} 
}
\startdata
$T_{\rm eff}$ (K) & 350 $\pm$ 10 &   357 $\pm$ 3  & 343 $\pm$ 11 & 386 $\pm$ 4 \\
log $g$ (cm s$^{-2}$) & 4.0 $\pm$ 0.2 & 4.2 $\pm$ 0.1   & 3.1 $\pm$ 0.1 &  4.42 $\pm$ 0.01 \\
Metallicity ([m/H] dex) & -0.25 $\pm$ 0.15 &  \nodata   & -0.13 $\pm$ 0.06 & -0.25 $\pm$ 0.04 \\
Age (Gyr) & \nodata & $1.6~^{+0.6}_{-0.4}$ &  \nodata  & 2.1 $\pm$ 0.6 \\
Luminosity ($L/L_{\odot}$) &  \nodata & -6.75 $\pm$ 0.02 &  \nodata  & -6.75 $\pm$ 0.01 \\
Radius ($R_{\rm Jupiter}$) &  1.11 $\pm$ 0.04  & 1.08 $\pm$ 0.02 & 1.12 $\pm$ 0.02 & 0.82 $\pm$ 0.02 \\
Mass ($M_{\rm Jupiter}$) &  \nodata & 7 $\pm$ 1 &  $0.75~^{+0.24}_{-0.16}$ & 7 $\pm$ 1 \\
\enddata
\end{deluxetable*}

\subsection{Bolometric Luminosity and Evolutionary Models}

To determine the bolometric luminosity of the Y dwarf, we first integrated the observed {\it JWST} spectra 
from Voy25 and Lew26 
across the wavelength range 2.88~$\mu$m to 11.99~$\mu$m. To estimate the flux contribution at shorter wavelengths we started with the ATMO 2020++ synthetic spectrum that best matched the observed   {\it JWST} F150W2 photometry  from Lew26 --- the 350~K log $g = 4.0$ [m/H] $= 0.0$ model --- and multiplied it by a factor of 2.15 to match the   F150W2 photometry (see inset figure in middle panel of Figure 1).  Although the factor of two discrepancy is concerning, the near-infrared energy distribution across the F150W2 filter will consist of three narrow flux peaks (Figure 1) and so the weighting with wavelength is reasonably well known; moreover  the $\lambda < 2.88~\mu$m region contributes $< 2\%$ of the total flux.  

The longer wavelength contributions were determined as follows. We used a synthetic spectrum scaled to match the F1500W photometry from Voy25 to determine the contribution at $11.99 < \lambda~\mu$m $< 30.0$; in this case the 350~K log $g = 4.0$ [m/H] $= -0.5$ model gave the best match and the spectrum was multiplied by a factor of 1.06 to match the   F1500W photometry (top panel of Figure 1). The contribution at $\lambda > 30~\mu$m was estimated to be a Rayleigh-Jeans function fixed at the scaled model spectrum at $\lambda = 30~\mu$m (where the ATMO 2020++ synthetic spectrum terminates).

Table 2 list the contributions from each wavelength region, and presents the total flux and luminosity.  The uncertainty for each observational section is equal to the uncertainty in the photometric calibration, 3\%.  We estimate an uncertainty of 5\% for the scaled model and Rayleigh-Jeans sections, based on tests using different models and different placements of the Rayleigh-Jeans curve. We arrive at the same luminosity value as Lew26 (see Table 1), although our approaches differ; Lew26 rely on the photometry and spectroscopy, and a Rayleigh-Jeans tail at $\lambda > 23~\mu$m.  

\begin{deluxetable*}{cccc}[!hb]
\tablecaption{Determination of Bolometric Luminosity for WD 0806-661B}
\tablehead{
\colhead{Wavelength} &   \colhead{Flux} & \colhead{\% of}  & \colhead{Method} \\
\colhead{Range $\mu$m} &  \colhead{W m$^{-2}$} & \colhead{Total} &
}
\startdata
0 -- 2.88 & 2.59 $\times 10^{-19}$ $\pm$ 5\% & 1.6 & ATMO 2020++ 350/4.0/0.0 scaled to F150W2 magnitude \\
2.88 -- 5.27 & 5.65 $\times 10^{-18}$ $\pm$ 3\% &  36.5 &  \citet{Lew_2026} NIRSpec spectrum \\
5.27 -- 11.99 & 3.89 $\times 10^{-18}$ $\pm$ 3\%  & 25.2 & \citet{Voyer_2025} MIRI spectrum \\
11.99 -- 30.00 & 5.11 $\times 10^{-18}$ $\pm$ 5\% & 33.0 & ATMO 2020++  350/4.0/-0.5 scaled to  F1500W magnitude \\
$>$ 30.00 & 5.76 $\times 10^{-19}$ $\pm$ 5\% & 3.7 & Rayleigh-Jeans tail \\ 
\enddata
\tablecomments{Total bolometric flux is 15.49 $\pm$ 0.79 $\times 10^{-18}$ W m$^{-2}$, 
at the BD distance of 19.23~pc. This translates to a luminosity of 6.85 $\pm$ 0.34 $\times 10^{19}$ W.   Adopting $L_{\odot} = 3.828 \times 10^{26}$ W, log $L/L_{\odot} = -6.75 \pm 0.02$.   }
\end{deluxetable*}

Table 1 gives the $T_{\rm eff}$, log $g$, $R$, and mass inferred from the evolutionary models by \citet{Marley_2021} using the luminosity and age of the system determined here.  The values for $T_{\rm eff}$ and log $g$ are in excellent agreement with the estimates based on the ATMO 2020++ grid comparison.  The values for $T_{\rm eff}$, log $g$, and $R$  lie between the values determined by Voy25 and Lew26 using retrieval methods on the MIRI data and NIRCam/NIRSpec data respectively (Table 1). 

\bigskip
\section{The Pressure-Temperature Curve of the Atmosphere of WD 0806-661B}

Figure 3 shows the pressure-temperature (P-T) relationships determined by the retrieval analyses of Voy25 and Lew26.  Also shown are the relationships for the metal-poor and solar-metallicity 350~K log $g = 4.0$ ATMO 2020++ atmospheric models adopted here.  Condensation curves for appropriate cloud-forming elements are also shown: ammonia and water at cold temperatures, and potassium chloride and sodium sulfide  at warmer temperatures \citep{Morley_2012, Morley_2014}. 
 Clouds would be expected to form where the condensation curves cross the P-T sequences.
Changes in the equilibrium nitrogen chemistry are also shown --- where N$_2$ is favored at higher temperatures and NH$_3$ at lower temperatures \citep[e.g.][their Figure 2]{Lodders_1999}.

\begin{figure}
\vskip -0.2in
\hskip 0.3in
\includegraphics[angle=-90, width = 6.8 in]
{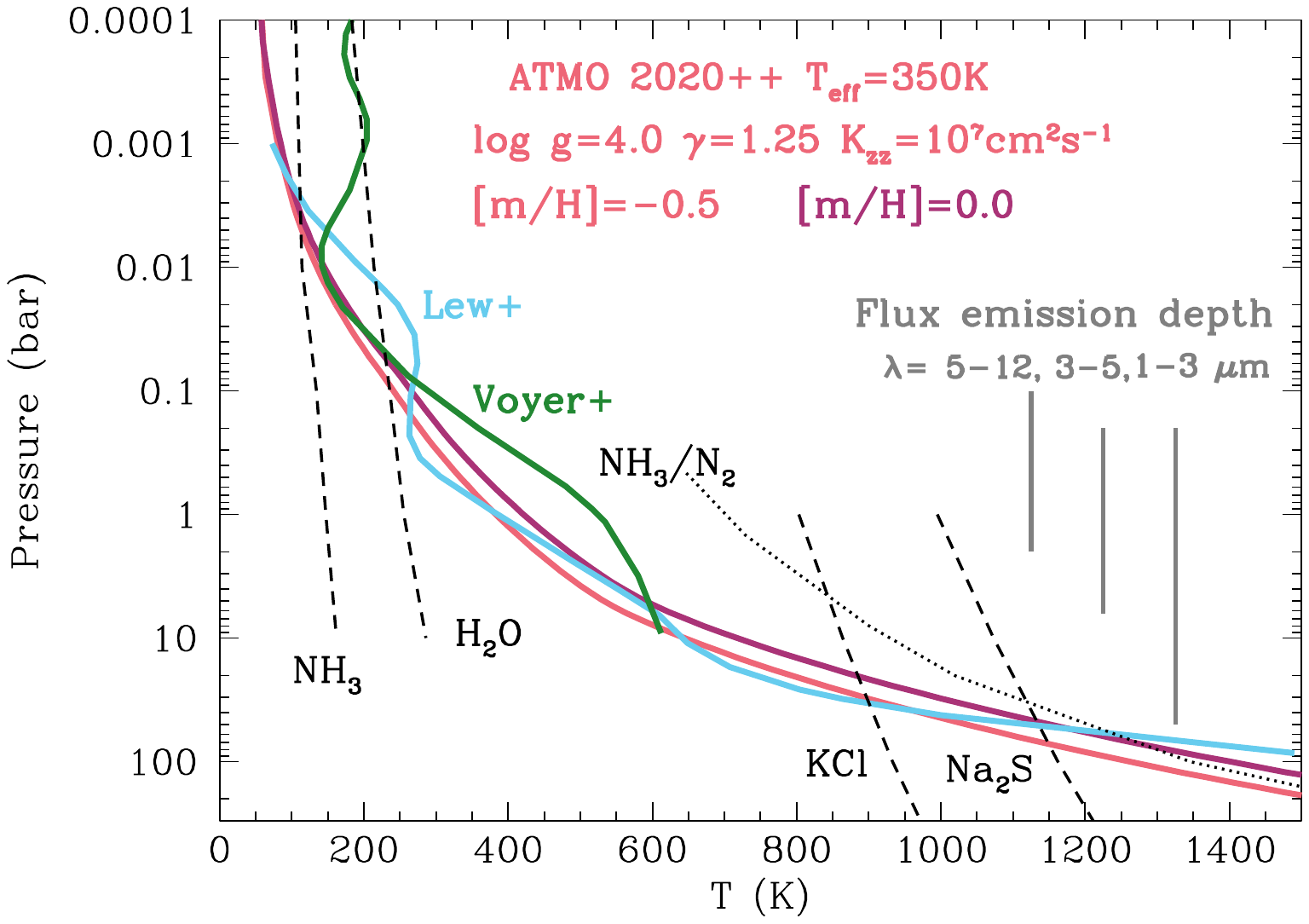}
\vskip -0.2in
\caption{Pressure-Temperature (P-T) relationships for the atmosphere of WD 0806-661B.  The green and cyan lines represent the relationships found by 
Voy25 and Lew26  respectively, using retrieval methods.  The pink and purple lines are those for the ATMO 2020++ models with parameters as shown in the legend.  The dashed black lines show the condensation curves for NH$_3$, H$_2$O, KCl, and Na$_2$S, as labelled;  the dotted line indicates where equilibrium nitrogen chemistry changes from N$_2$-dominant at higher temperatures to NH$_3$-dominant at lower temperatures.
Dark gray lines indicate the depths from which the near-infrared, NIRSpec and MIRI light emerges. 
}
\end{figure}

Vertical gray lines in Figure 3 indicate the contribution functions determined by the retrieval analyses 
(Lew26, their Figure 5,  and Voy25, their Figure 4)  and the ATMO 2020++ models \citep[][their Figure 7]{Leggett_2021}.  These demonstrate the pressures (depths) from which light of a certain wavelength typically originates.  

The short-wavelength near-infrared light, sampled by the Lew26 NIRCam photometry, arises from 
the deeper regions of the atmosphere, at pressures of tens of bar. 
 Here 
the retrieved profile by Lew26 is similar to our preferred ATMO 2020++ model profiles, although 
it is likely that the P-T curves in this region should be warmer in order to increase the emergent flux and better match the observations (inset figures in Figure 1). The P-T slope at pressures of tens of bar is subadiabatic, as found empirically for the ATMO 2020++ suite, and as determined by the Lew26 retrieval. 
At these depths condensation of KCl and Na$_2$S and the shift in the nitrogen equilibrium chemistry are important processes (Figure 3) which may disrupt convection \citep{Tremblin_2015, Tremblin_2019,Tan_2017}.

The energy at the wavelengths sampled by the Voy25 MIRI data arises from higher and colder layers of the atmosphere, at pressures of tenths of a bar. Here the Lew26 P-T curve is colder than the others, which leads to the too-faint spectrum calculated by Lew26 at these wavelengths.   In the Voy25 retrieval of the MIRI data, the contribution function shows a strong peak around 0.2~bar where $T \approx 360~$K. In that study, the pressure scale of the atmosphere is controlled by mass, and those authors find  an exceedingly low mass for the BD, of $< 1 ~M_{\rm Jupiter}$.   Voy25 suggest that the retrieved value of the mass could be biassed by 
assumptions made regarding the opacity treatment, or the use of free chemical abundances.

Voy25 find no evidence of H$_2$O clouds. Although both the Voy25 and the ATMO 2020++ P-T curves intersect the H$_2$O condensation line, these very low pressures are probed by wavelengths where there is very little signal:  at 6 -- 8~$\mu$m, and at 10.5~$\mu$m at the bottom of the strong NH$_3$ absorption  \citep[Figure 7 of][and Figure 5 of Lew26]{Leggett_2021}.  Constraining the presence or absence of H$_2$O clouds using these data is therefore difficult.

\bigskip
\section{Conclusions}

Voy25 and Lew26 present high signal-to-noise {\it JWST} data for a faint but important source --- the planetary-mass companion to the white dwarf WD 0806-661.   Each team studies different datasets with different wavelength coverage --- either the NIRSpec data or the MIRI data.  Both teams allow the radius of the BD to float, and not be constrained by evolutionary models.  Both analyses produce results which are inconsistent with brown dwarf evolutionary theory.

Here we study both datasets together, for increased wavelength coverage.  We also take a more pragmatic approach where we assume that the evolutionary models are more robust than the atmospheric models, and we constrain the radius of the BD to be the value calculated by the evolutionary models for the $T_{\rm eff}$ and log $g$ of each atmospheric model.  
We estimate the physical parameters of the BD using two approaches.  One is a comparison to ATMO 2020++ synthetic spectra, and the other  uses the bolometric luminosity of the BD.  

The ATMO 2020++ models have been empirically adjusted to better reproduce the observed energy distributions of Y dwarfs.  They include a diabatic treatment of the P-T curve at pressures around 10 bar, which may be due to kinetic or chemical processes. This region is probed by the  NIRCam and NIRSpec data, and here the adopted ATMO 2020++ P-T curves are similar to those of Lew26.  However the floating radius used by Lew26 leads to a warmer effective temperature and higher gravity (Table 1). The ATMO 2020++ models that we adopt here indicate an   effective temperature similar to that found by Voy25, however the gravity and mass that we determine are very different from those of Voy25 (Table 1).

We determine a value for the bolometric luminosity equal to the value found by Lew26.  Using a slightly revised age for the system, constrained by the white dwarf primary, we determine an  effective temperature and surface gravity using evolutionary models that is in excellent agreement with the values determined by  our ATMO 2020++ comparison. These parameters are given in Table 1.

Until we better understand the turbulent and planet-like atmospheres of the Y dwarfs, we recommend that evolutionary models are used to constrain analyses of data.  Also, analyses need to use observations with the broadest wavelength coverage possible, near- to mid-infrared, to determine reliable atmospheric properties for these cold objects.  Allowing properties such as radius to float, or using datasets with only near-infrared or only mid-infrared coverage, can lead to results which are artificially pushed to implausible values.

\bigskip

\begin{acknowledgments}

We are grateful to Ben Lew and Mael Voyer for sharing their reduced data with us.

This work is based on observations made with the NASA/ESA/CSA {\it James Webb Space Telescope}. The data were obtained from the Mikulski Archive for Space Telescopes at the Space Telescope Science Institute, which is operated by the Association of Universities for Research in Astronomy, Inc., under NASA contract NAS 5-03127 for JWST. These observations are associated with GTO program 1276. The specific observations can be accessed via \dataset[doi:10.17909/anme-j453]{\doi{10.17909/anme-j453}}
and \dataset[doi:10.17909/dacd-v990]{\doi{10.17909/dacd-v990}}.
\end{acknowledgments}

\bigskip
\bibliography{0806}{}
\bibliographystyle{aasjournal}

\end{document}